\documentclass[12pt]{article}
\usepackage[english]{babel}
\usepackage[pdftex]{graphicx}
\usepackage[authoryear,round]{natbib}
\usepackage[margin=20mm]{geometry}
\usepackage{hyperref}
\hypersetup{
    colorlinks = true,
    linkcolor = blue,
    anchorcolor = blue,
    citecolor = blue,
    filecolor = blue,
    urlcolor = blue,
	pdfauthor={some author},
	pdftitle={eye-catching title}
    }
\usepackage{amssymb,amsmath}
\usepackage{url}

\begin{document}

\title{The jetted NLS1 1H~$0323+342$: the Rosetta stone for accretion/ejection in AGN\footnote{Presented at: \emph{Highly accreting supermassive black holes across all cosmic times: from the local Universe to cosmic dawn}, December 1-5, 2025, European Southern Observatory, Vitacura, Santiago (Chile).}}
\author{Luigi Foschini\footnote{Brera Astronomical Observatory, National Institute of Astrophysics (INAF), Milano/Merate, Italy}.}
\date{November 30, 2025}
\maketitle

\begin{abstract}
1H~$0323+342$ is the nearest gamma-ray narrow-line Seyfert 1 galaxy ($z=0.063$). Its X-ray spectrum ($0.3-10$~keV) is characterised by significant spectral variability observed by many authors, with a backbone with photon index $\sim 2$ occasionally superimposed by a hard tail. This spectral variability has been interpreted as the interplay between the X-ray corona and the relativistic jet. The X-ray fluxes in the $0.3-10$~keV energy band are generally around $\sim 10^{-11}$~erg~cm$^{-2}$~s$^{-1}$, making it easier to get sufficient statistics even with short exposures. Here I present a reanalysis of all the available X-ray observations with \textit{Swift} (181 obs), \textit{XMM-Newton} (7 obs), \textit{Chandra} (1 obs), and \textit{Suzaku} (2 obs) performed between 2006 and 2025. Possible interpretations are proposed and discussed. 
\end{abstract}

\section{Introduction}
Understanding the accretion/ejection engine in active galactic nuclei (AGN) is an evergreen conundrum. Obviously, after decades of research, we know that these two processes are linked together, there are valuable theories, but devil is in the details. In the present work, I would like to push the devil out of these details, at least for one case study. 

The $\gamma-$ray narrow-line Seyfert 1 (NLS1) galaxy 1H~$0323+342$ ($z=0.063$, \citealt{ZHOU2007}) offers an interesting opportunity. It is the closest $\gamma-$ray NLS1, with a relatively small mass of the central black hole ($\sim 2\times 10^{7}M_{\odot}$) and strong accretion disk close to the Eddington limit \citep{LANDT2017}, whose emission extends in the X-ray energy band. Also the powerful relativistic jet can give a significant contribution to the X-ray emission. The resulting X-ray flux in the $0.3-10$~keV energy band exceeds almost always the value of $\sim 10^{-11}$~erg~cm$^{-2}$~s$^{-1}$, making it possible to get a rough measurement of flux and photon index even with shallow observations. This means that 1H~$0323+342$ offers the almost unique possibility for a jetted AGN to explore the accretion/ejection processes by X-ray observations, like X-ray binaries.

The different contributions to the X-ray emission can be disentangled by looking at the X-ray photon index. The thermal disk emission -- a multicolor black body\footnote{By assuming a standard disk; however, given the high accretion rate, it would be interesting to explore the possibility of slim disk models.} -- can be represented also with a power-law model with a very soft photon index, while the X-ray corona is the result of the unsaturated Comptonization and can be modeled with a power-law with photon index around $\sim 1.9$. The jet emission in the X-ray energy band is mostly due to the scattering of relativistic electrons off the broad-line region (BLR) photons (cf. \citealt{LAT2009}), which can be modeled again with a power-law, but with a hard photon index ($<<1.9$).

\begin{figure}[ht]
\begin{center}
\includegraphics[scale=0.4]{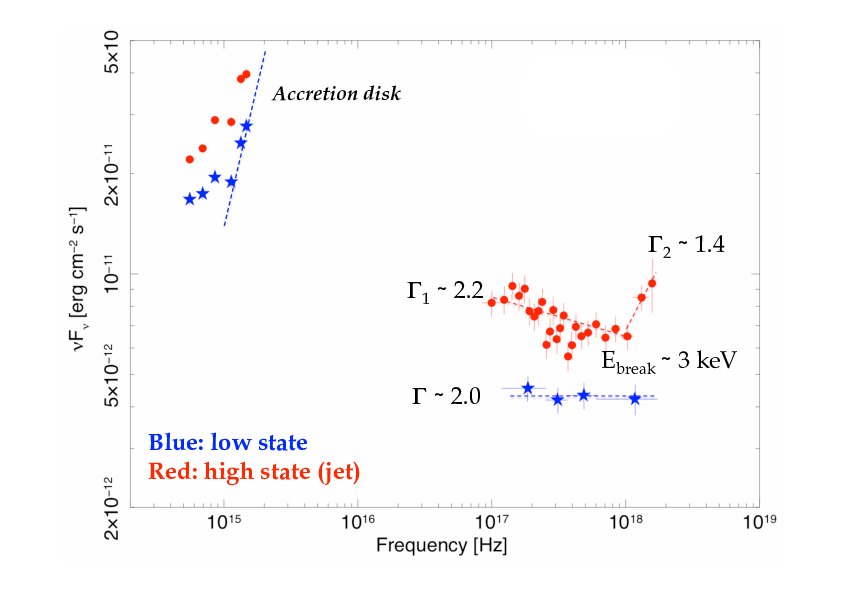}
\caption{Spectral changes as observed by \emph{Swift} satellite (from \citealt{FOSCHINI2012}).}
\label{xrtbpl}
\end{center}
\end{figure}

Early evidence of such strong spectral changes was provided by \cite{FOSCHINI2009}, who reported soft and low flux as observed by \emph{INTEGRAL}/ISGRI, but hard and high flux with \emph{Swift}/BAT observation in the $20-100$~keV energy band. Also \emph{Swift}/XRT provided evidence of spectral changes in the $0.3-10$~keV band \citep{FOSCHINI2012}: a corona dominated state with $\Gamma\sim 2.0$ and a jet dominated state, with a spectral break around $\sim 3$~keV and a hard photon index $\sim 1.4$ (Fig.~\ref{xrtbpl}).  

\begin{figure}[ht]
\begin{center}
\includegraphics[scale=0.4]{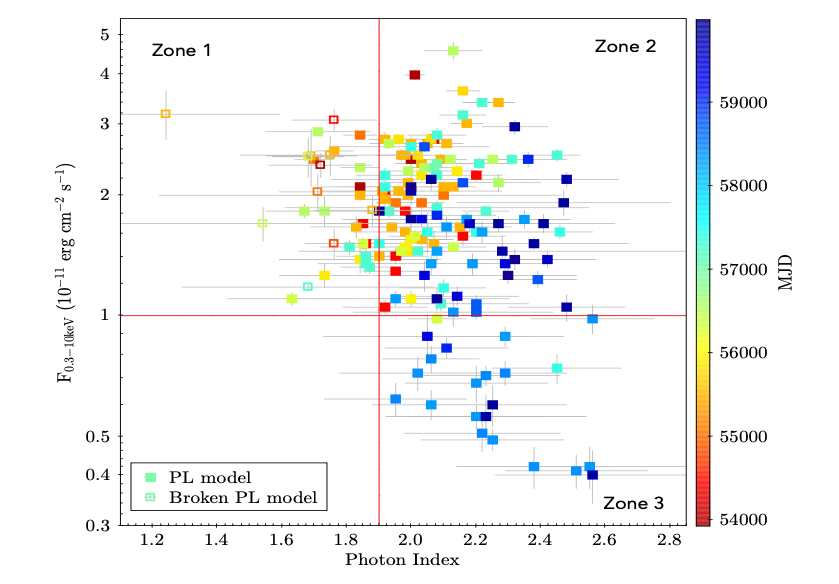}
\caption{Total X-ray flux ($0.3-10$~keV) as a function of the photon index as observed by \emph{Swift} satellite (from \citealt{ROSA2025}).}
\label{rosa}
\end{center}
\end{figure}

These spectral changes were confirmed by \cite{TIBOLLA2013,PALIYA2014,FOSCHINI2015,YAO2023,ROSA2025}. Particularly interesting are the finding of \cite{ROSA2025}, who analysed all the 172 \emph{Swift} observations from 2006 to 2023, and found that the X-ray emission can be divided roughly into three zones, depending on the combination of the total flux and photon index (Fig.~\ref{rosa}): 

\begin{enumerate}
\item when the $F_{0.3-10\,\rm{keV}}\gtrsim 10^{-11}$~erg~cm$^{-2}$~s$^{-1}$ and $\Gamma \lesssim 1.9$, which is likely to be dominated by the jet;
\item when the $F_{0.3-10\,\rm{keV}}\gtrsim 10^{-11}$~erg~cm$^{-2}$~s$^{-1}$ and $\Gamma \gtrsim 1.9$;
\item when the $F_{0.3-10\,\rm{keV}}\lesssim 10^{-11}$~erg~cm$^{-2}$~s$^{-1}$ and $\Gamma \lesssim 1.9$, which is likely to be dominated by the X-ray corona, while the jet is almost inactive.
\end{enumerate}

\cite{ROSA2025} proposed that the observed patterns could be explained in term of intermittent jet activity due to accretion disk instabilities according to the theory by \cite{CZERNY2009}.

A more detailed spectral analysis of \emph{XMM-Newton} observations by \cite{MEHDIPOUR2019}, revealed the presence of disk winds with column densities of $(9\pm2)$ and $(7.2\pm0.7)\times 10^{20}$~cm$^{2}$, with ionization parameter $\log \xi\sim 2.17,$ and $0.15$, and outflow velocity of $\sim 830$, and $880$~km/s, respectively. From an optical point of view, \cite{BERTON2016} found that the [OIII]$\lambda 50007$ emission line core is not blue- or red-shifted within the measurement error, but it displays a significant blue wing with full-width half-maximum (FWHM) equal to $\sim 1200$~km/s (to be compared with the FWHM of the core component $\sim 445$~km/s).

Another piece of the puzzle comes from the study of other NLS1s. \cite{LAHTEENMAKI2018} reported the discovery of significant radio emission (at Jy level) at 37~GHz from radio-quiet or even silent NLS1s. Follow-up at radio frequencies by \cite{BERTON2020,JARVELA2021} suggested the existence of free-free absorption, which hides the radio emission at low frequencies ($\lesssim 10$~GHz). Hence the name ``absorbed jets'' \cite{BERTON2020}, or ``weirdos'' as Emilia J\"arvel\"a called them at the workshop ``Observations and physics of NLS1 galaxies: AGN at their extreme'' (ESO, Santiago, Chile, 26-28 November 2025). Since these outbursts are rather erratic, an optical/UV/X-ray monitoring was set up with the \emph{Swift} satellite targeting the most promising source among those identified, SDSS~J$164100.10+345452.7$. Expectations were not disappointed and \cite{ROMANO2023} reported an X-ray spectral change close to the radio outburst. When the source is undetected at radio frequencies, the X-ray spectrum shows a strong intrinsic absorption, which disappears close to the radio burst. The timescales seem to be quite short, around hours. 

\begin{figure}[ht]
\begin{center}
\includegraphics[scale=0.2]{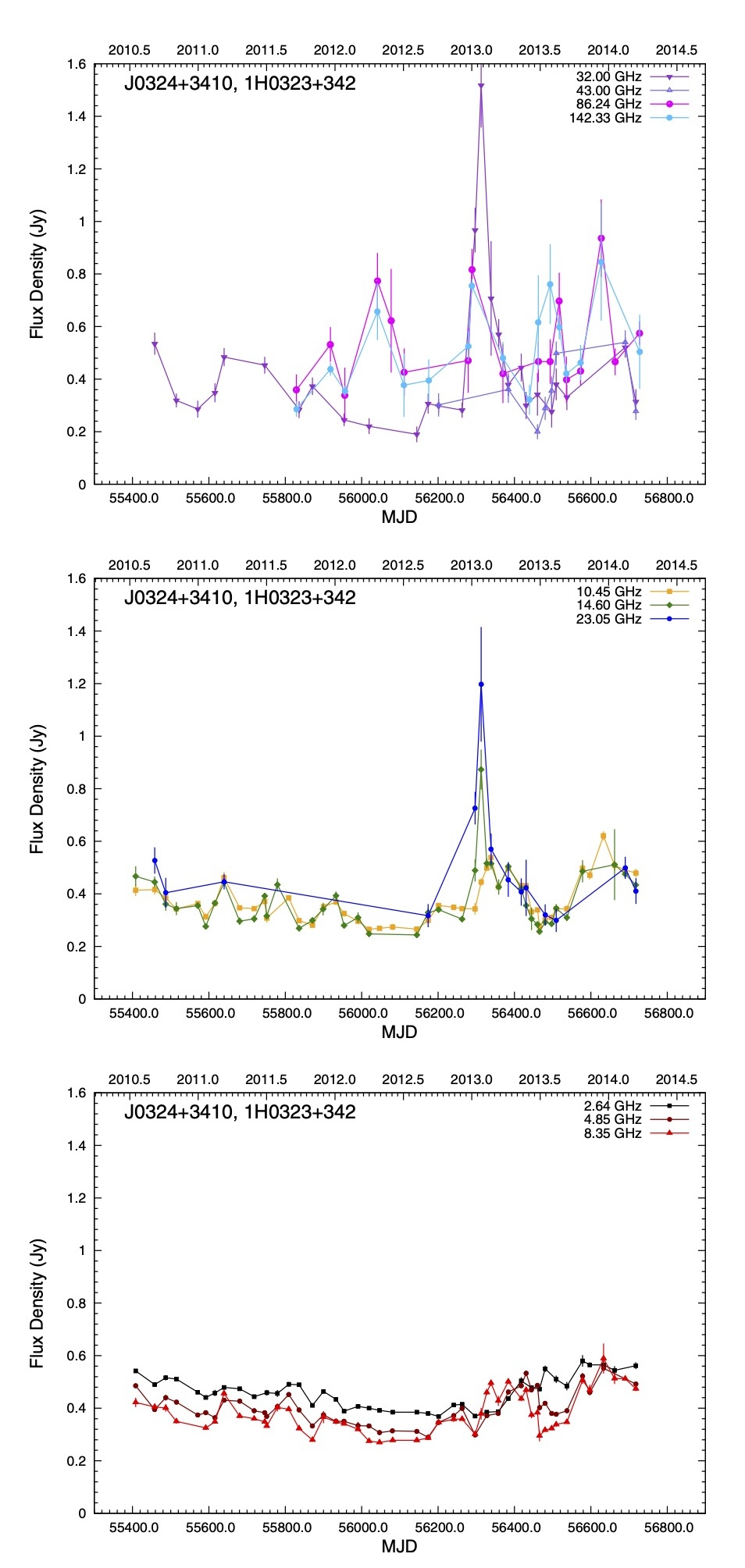}
\caption{Effelsberg multifrequency radio observations of 1H~$0323+342$ (from \citealt{ANGELAKIS2015}).}
\label{effelsberg}
\end{center}
\end{figure}

It is worth comparing the radio emission of the above cited NLS1s and 1H~$0323+342$: \cite{ANGELAKIS2015} reported about the long-term monitoring of a sample of $\gamma-$ray emitting NLS1s, including 1H~$0323+342$. They found that for frequencies $\lesssim 10$~GHz, the radio emission is almost stable, while strong variability appears at higher frequencies, with observations at $14.60$~GHz displaying the largest amplitudes (Fig.~\ref{effelsberg}). The flare on MJD $56313$ peaked at $32$~GHz. The fact that 1H~$0323+342$ is detected at low frequencies, while L\"ahteenm\"aki's sample is not, may be due to different optical depth of the warm absorber. 

To summarize, 1H~$0323+342$ displays strong spectral variability at X-rays, outflows/winds at optical and X-rays, a radio behavior somehow similar to ``weirdos''. Therefore, I have decided to analyse/reanalyse all the available X-ray observations, trying to understand how accretion and ejection processes are linked with outflows and warm absorption. I report here some preliminary result, because the work is still ongoing. 

In the following, I adopt the reference values of 1H~$0323+342$ as listed in \cite{FOSCHINI2019}. Luminosities have been calculated by using $H_0=73.3$~km~s$^{-1}$~Mpc$^{-1}$ \citep{RIESS2022}. 

\section{Data Selection and Analysis}

\subsection{Chandra}
The \emph{Chandra} satellite observed the object only once, on November 8th, 2021, with about $10$~ks exposure. Data of the obsID $26173$ were retrieved from the \texttt{Chandra X-ray Center}\footnote{\url{https://cda.harvard.edu/chaser/}}, and reprocessed by using \texttt{CIAO v. 4.17} and \texttt{CALDB v. 4.12.0}. Given the fluxes of the order of $\sim 10^{-11}$~erg~cm$^{-2}$~s$^{-1}$, one can expect pile-up, but this was mitigated by placing the source off-axis, on the border of the chip ASICS-S3. A further mitigation can be obtained by selecting only grade 0 (single) and status 0 events. Therefore, I adopted an circular extraction region with radius of $6''$, while a radius of $120''$ was adopted for the background. The $10$~ks observation was then divided into four bins of $2.5$~ks elapsed time each. The extracted spectra were rebinned to have a minimum of 20 counts per bin, to apply the $\chi^2$ statistics, and fitted in the $0.3-7$~keV energy range, while the flux was extrapolated to the $0.3-10$~keV energy band for comparison with the observations of other satellites.

\subsection{Suzaku}
\emph{Suzaku} observed 1H~$0323+342$ twice, one on July 26th, 2009 (obsID $704034010$, exposure $74$~ks in $3\times3$ mode, and $9.7$~ks in $5\times5$ mode), and the other on March 1st, 2013 (obsID $70715010$, exposure $87$~ks in $3\times3$ mode, and $15$~ks in $5\times5$ mode). XIS0, XIS1, and XIS3 data were downloaded from \texttt{HEASARC} website\footnote{\url{https://heasarc.gsfc.nasa.gov}}, reprocessed by using \texttt{HEASoft v. 6.35.1} and \texttt{CALDB} updated on July 14th, 2025. Standard procedures were used (see \emph{The Suzaku Data Reduction Guide}, v. 5.0, available on \texttt{HEASARC}). The source extraction region was a circle with radius $260''$, while the background was extracted from a rectangular region with the maximum source-free area, taking into account the source position and the geometry of the chip (the resulting areas were from $~2.8\times 10^5$~arcsec$^2$ to $~4.2\times 10^5$~arcsec$^2$). The extracted spectra (in the $0.3-10$~keV energy band for XIS1, and $0.6-10$~keV band for XIS0 and XIS3) were rebinned to have at least 20 counts per bin and fitted simultaneously by using XIS0 as reference. The observations were divided into one-hour elapsed time bins, with some difference in the last bin to match the total elapsed time. Therefore, the effective exposures can vary. The total number of bins was 48 for the obsID $704034010$, and 52 for obsID $70715010$.

\subsection{Swift}
The \emph{Swift} satellite observed the object 182 times from July 6th, 2006, to October 17th, 2025. All the observations were downloaded from \texttt{HEASARC} website\footnote{\url{https://heasarc.gsfc.nasa.gov}}, reprocessed and analysed by using \texttt{HEASoft v. 6.35.2} and \texttt{CALDB} updated on October 23th, 2025. Standard procedures were used (see \emph{The Swift XRT Data Reduction Guide}, v. 1.2, available on \texttt{HEASARC}). Data of the \emph{X-Ray Telescope} (XRT) were available both in photon counting (PC) and window timing (WT) mode, but almost all the exposure was for the former, while the latter had only a few tens of seconds. Therefore, I analysed only PC data from pointed observations, which resulted to be 181 (the obsID~03111698008 contained only slew and settling data, and therefore was discarded). Exposures range from $\sim 0.1$ to $\sim 9$~ks, but even in the worst case there were counts enough (obsID~00096107007, exposure $\sim 122$~s, $26$~cts) to get a rough spectrum to be fit with a power-law model. The extracted spectra were rebinned to ensure at least 20 counts per bin, to apply the $\chi^2$ statistics. When the statistics is not enough, the likelihood was used \citep{CASH1979}. 

Since the expected fluxes are of the order of $\sim 10^{-11}$~erg~cm$^{-2}$~s$^{-1}$, I performed some check (circular or annular extraction region, grades including only single or multiple events) to understand if the point-spread function (PSF) can be affected by pile-up. Therefore, for the sake of simplicity and given the large number of observations, I selected an annular extraction region. Since the maximum count rate is $\sim 1$~c/s (obsID 00035372003, on September 21st, 2025, during the follow-up of the $\gamma-$ray outburst reported by \citealt{LONGO2025}), I adopted the prescription by \cite{EVANS2014} and selected the inner radius of $\sim 10''$ and the outer radius of $47''$. 

\subsection{XMM-Newton}
Seven \emph{XMM-Newton} observations were found in the science archive\footnote{\url{https://nxsa.esac.esa.int/nxsa-web/\#search}}, retrieved, and reprocessed by using \texttt{Science Analysis Software SAS v. 22.1} with calibration files \texttt{CCF} updated on March 17th, 2025. The obsIDs are: $076460101$ (exposure $\sim 64$~ks, taking the EPIC-PN detector as reference), $0823780201$ (exposure $\sim 45$~ks), $0823780301$ (exposure $\sim 41$~ks), $0823780401$ (exposure $\sim 41$~ks), $0823780501$ (exposure $\sim 46$~ks), $0823780601$ (exposure $\sim 45$~ks), $0823780701$ (exposure $\sim 47$~ks). The data were reprocessed and cleaned for soft-proton flares by using the standard threads available on the \emph{XMM-Newton} web site\footnote{\url{https://www.cosmos.esa.int/web/xmm-newton/how-to-use-sas}}.

Also in this case, the long exposures were divided into one-hour elapsed time bins: 22 bins for obsID $076460101$, 15 for obsID $0823780201$, and 14 for all the remaining. The two MOS detectors were operated in Small Window mode, so that they did not suffer of pile-up problems: the extraction regions were circles with $40''$ radius for both the source and the background. The PN was set in Large Window mode, requiring an annular extraction region (inner radius $20''$, outer radius $40''$) to minimize the pile-up. To further mitigate the pile-up, I adopted \texttt{PATTERN=0}, i.e. I selected only single events. 

\subsection{Analysis}
From the above strategy, I obtained 392 spectra. Spectra were analysed by using the \texttt{xspec} software package. Since the aim of this work is to study the spectral changes in time, I adopted simple models: power law (\texttt{zpowerlw}) with photon index $\Gamma$, broken power law (\texttt{zbknpower}), with $\Gamma_1$ and $\Gamma_2$, photon indexes below and above the break energy $E_{\rm break}$ [keV], respectively, and absorption edge (\texttt{zedge}), with energy $E_{\rm edge}$ [keV] and optical depth $\tau$. The adopted redshift is $0.063$ \citep{ZHOU2007}. The Galactic hydrogen column was fixed to the value of $N_{\rm H}=1.17\times 10^{21}$~cm$^{-2}$ \citep{HI4PI} in the T\"ubingen-Boulder InterStellar Matter absorption model (\texttt{tbabs}, \citealt{WILMS2000}). The threshold for preferring one model over another is that the fit has to improve at 99\% confidence level (cf \citealt{PROTASSOV2002}). 

\begin{figure}[!ht]
\begin{center}
\includegraphics[scale=0.3]{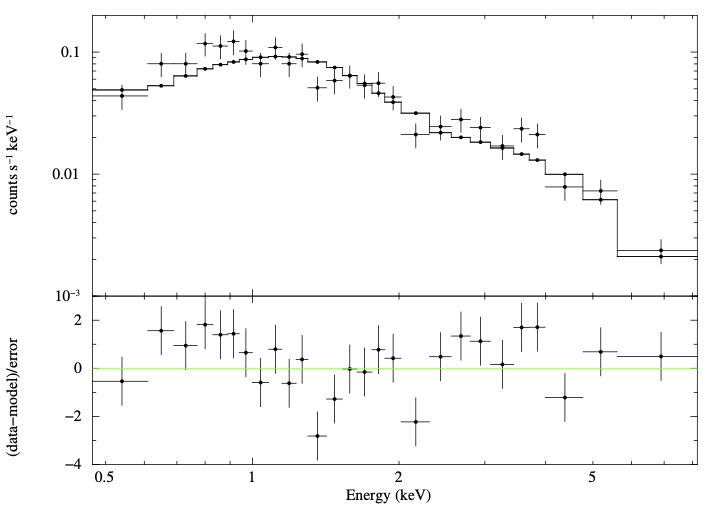}\\
\includegraphics[scale=0.3]{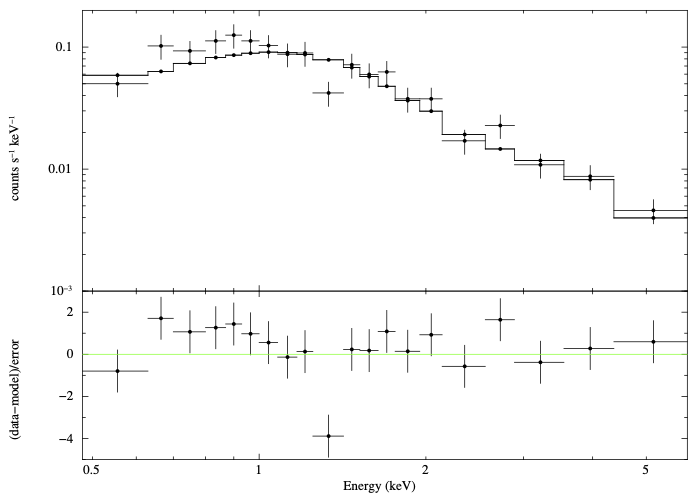}\\
\includegraphics[scale=0.3]{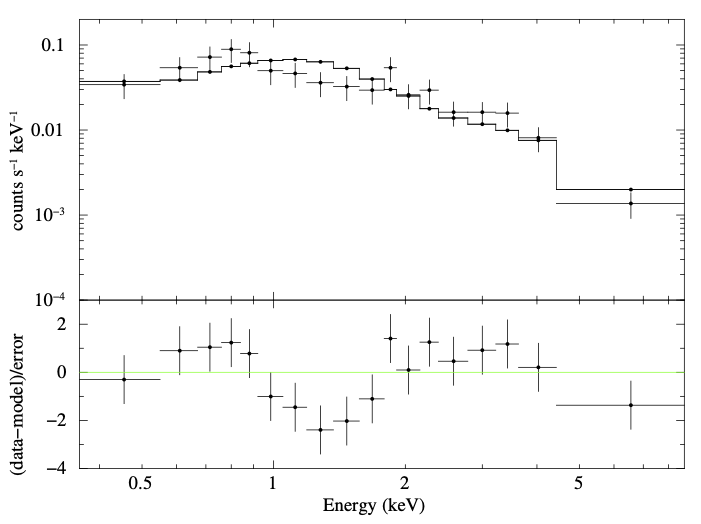}
\caption{The examples of \emph{Swift} spectra fitted with a power-law model, showing clear deviations then modeled with an absorption edge. (\emph{top panel}) ObsID $00036533013$, July 24, 2009; in this case, only the addition of the absorption edge at $1.35_{-0.15}^{+0.08}$~keV and $\tau=0.63_{-0.32}^{+0.37}$ was significant. (\emph{middle panel}) ObsID $00036533014$, July 27, 2009; absorption at $1.29_{-0.12}^{+0.07}$~keV with $\tau=0.72_{-0.35}^{+0.41}$. (\emph{bottom panel}) ObsID $00036533050$, September 27, 2013; broad absorption edge at $1.05_{-0.07}^{+0.08}$~keV and $\tau=1.38_{-0.69}^{+0.81}$. }
\label{zedge}
\end{center}
\end{figure}

\section{Results}
The first important result is to confirm warm absorption in $44$ of the $392$ spectra. Fig.~\ref{zedge} shows a few example of spectra with evident absorption edge. It is worth noting that almost simultaneously to these first two \emph{Swift} observations there was also one by \emph{Suzaku} (obsID $704034010$, July 26-28, 2009), which measured the edge at $1.32\pm 0.04$~keV, consistent with the \emph{Swift} measurements, and with $\tau=0.46_{-0.14}^{+0.15}$. The edge detected by \emph{Suzaku} was observed after $\sim 2.7$~days the first \emph{Swift} detection, and precedes the second \emph{Swift} detection by only half a day ($0.51$~d). \emph{XMM-Newton} was able to catch changes  on hourly timescales in August 2015 and August 2018 observations. The detection of absorption features with different instruments strengthen the significance of the result. 

\begin{figure}[!t]
\begin{center}
\includegraphics[scale=0.3]{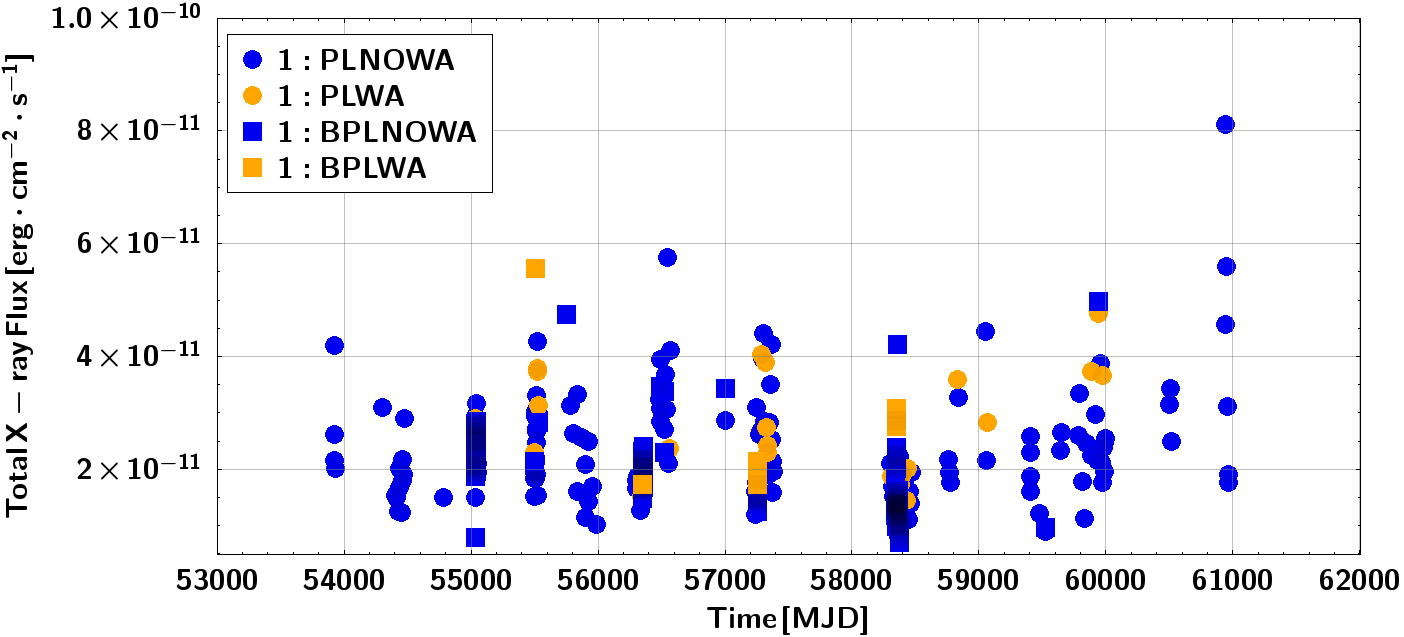}
\caption{Total X-ray flux in the $0.3-10$~keV energy band vs time. Label meaning: PLNOWA, power-law model with no edge; PLWA, power-law model with absorption edge; BPLNOWA, broken power-law model with no edge; BPLWA, broken power-law model with absorption edge.}
\label{lcurve}
\end{center}
\end{figure}

\begin{figure}[!t]
\begin{center}
\includegraphics[scale=0.3]{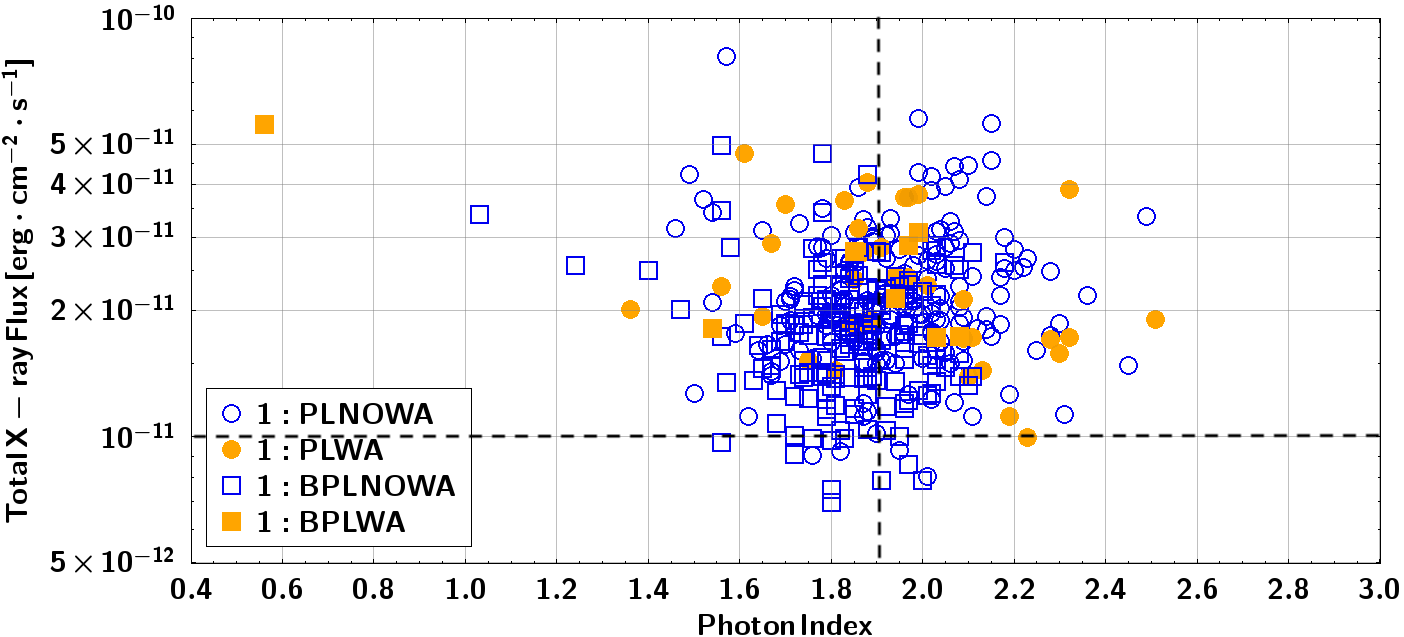}
\caption{Photon Index vs total X-ray flux in the $0.3-10$~keV energy band. The dashed lines are the references in the Rosa's plot (Fig.~\ref{rosa}, \citealt{ROSA2025}). Label meaning: PLNOWA, power-law model with no edge; PLWA, power-law model with absorption edge; BPLNOWA, broken power-law model with no edge; BPLWA, broken power-law model with absorption edge.}
\label{gammavsflux}
\end{center}
\end{figure}

\begin{figure}[!ht]
\begin{center}
\includegraphics[scale=0.3]{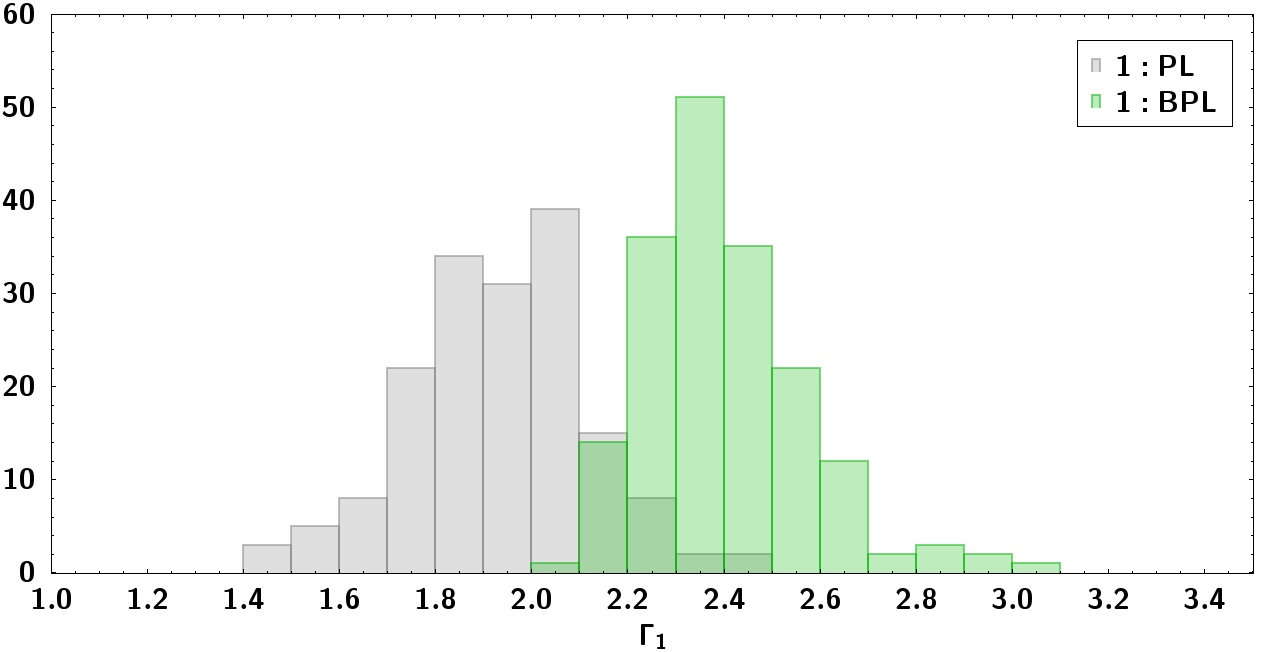}\\
\includegraphics[scale=0.3]{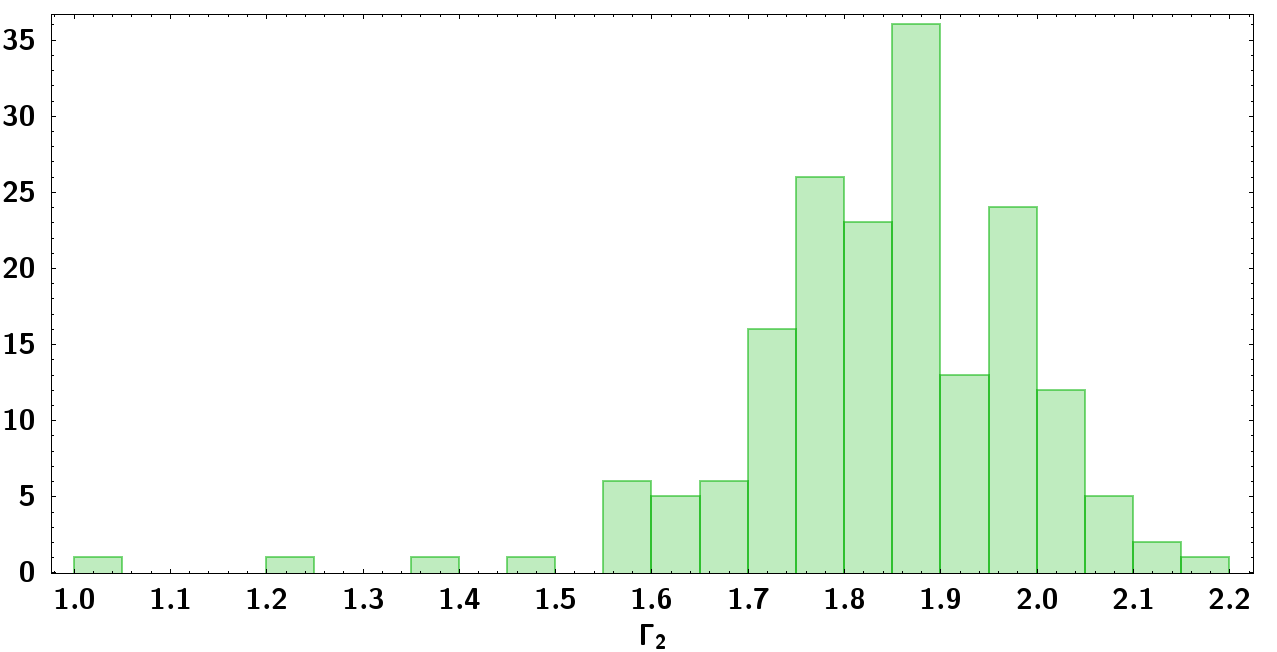}\\
\includegraphics[scale=0.3]{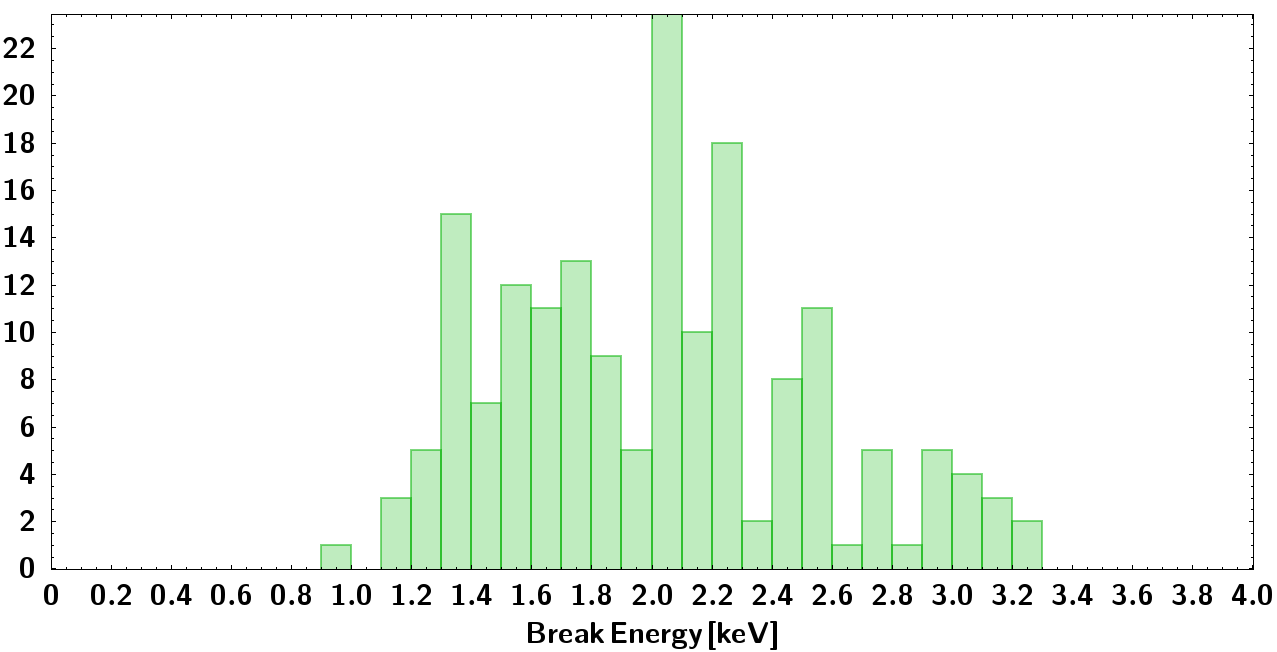}
\caption{(\emph{top panel}) Distribution of the photon indexes of power-law model ($\Gamma$) together with the $\Gamma_1$ of the broken power-law model. (\emph{middle panel}) Distribution of the $\Gamma_2$ of the broken power-law model. (\emph{bottom panel}) Distribution of the $E_{\rm break}$ [keV] of the broken power-law model.}
\label{gammadistr}
\end{center}
\end{figure}

\begin{figure}[!ht]
\begin{center}
\includegraphics[scale=0.3]{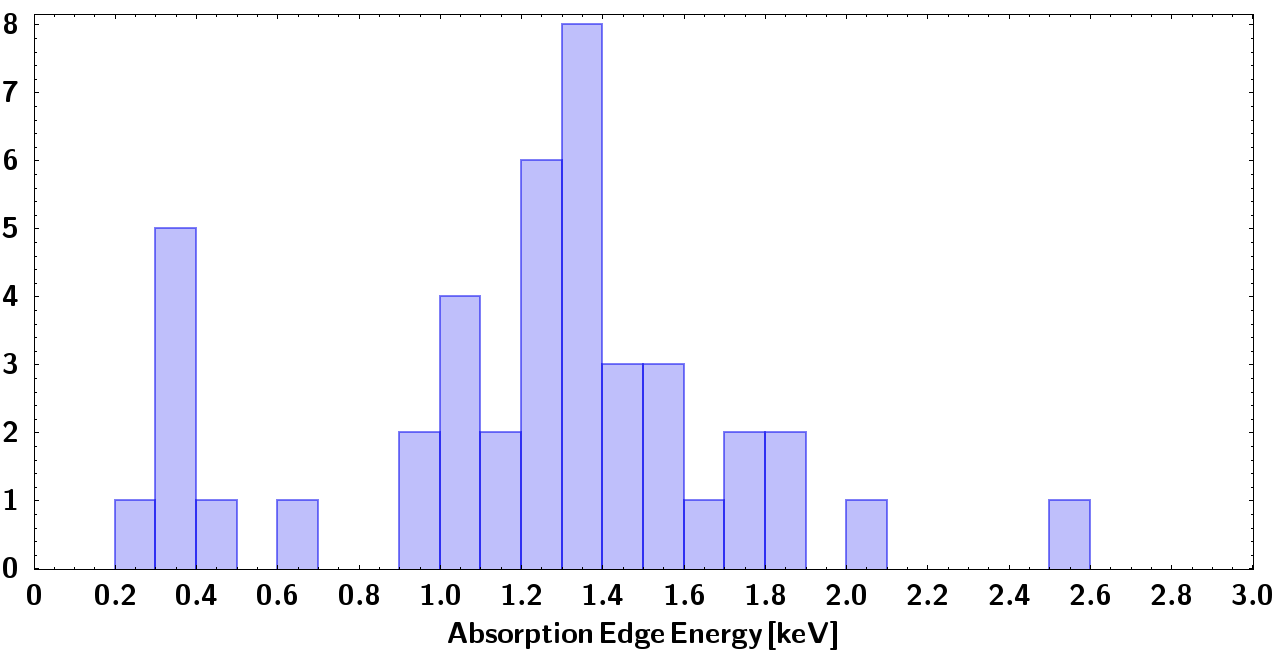}\\
\includegraphics[scale=0.3]{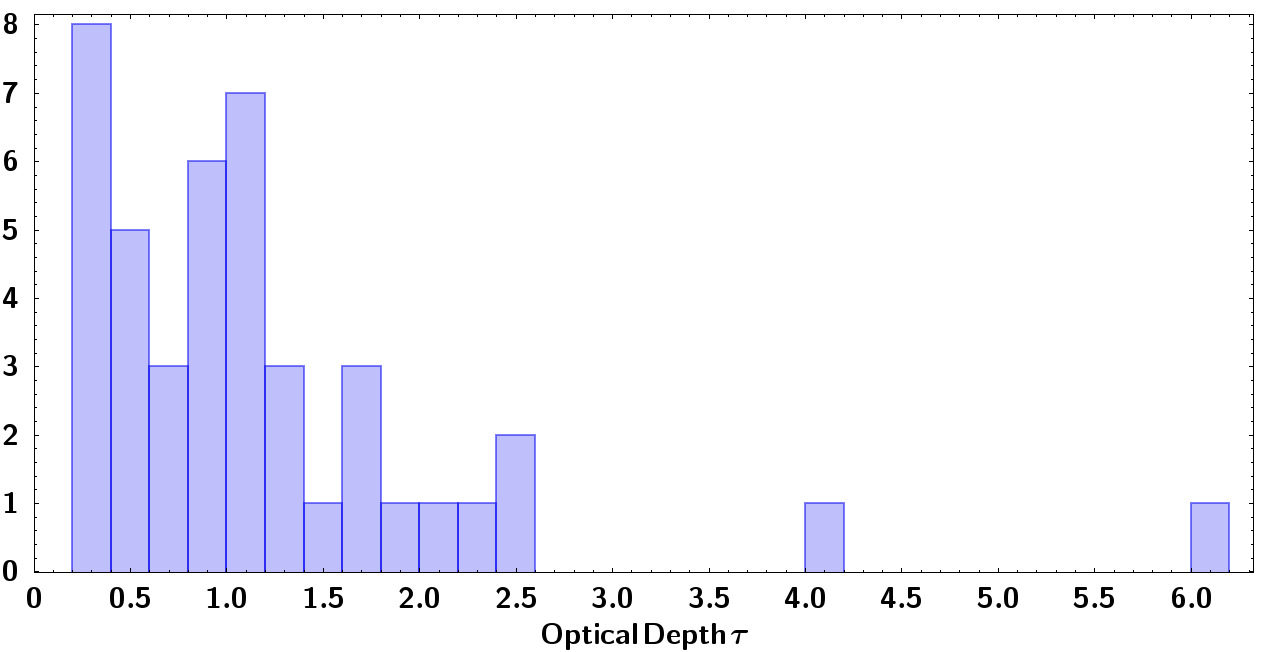}
\caption{(\emph{top panel}) Distribution of the rest frame energy of the absorption edges $E_{\rm edge}$ [keV]. (\emph{bottom panel}) Distribution of the optical depths. Please note that the two extreme values 4 and 6 are upper limits.}
\label{edgesdistrib}
\end{center}
\end{figure}

Fig.\ref{lcurve} shows the lightcurve obtained by using all the 392 spectra. A first look suggests that the warm absorbers are not always present, but appear under certain circumstances to be understood. The photon index--flux plot (Fig.~\ref{gammavsflux}) does not suggest any trend, with the absorption edges randomly distributed either with soft or hard spectra. Also in this case, I considered the total X-ray flux and $\Gamma_{2}$ (the photon index for energies above the break) to represent the broken power-law model. I also noted that now the photon index--flux plot can be basically divided into two zones only, by keeping the photon index of the unsaturated Comptonization as reference: the jet zone, with $\Gamma \lesssim 1.9$ and the corona zone, with $\Gamma \gtrsim 1.9$.

Fig.~\ref{gammadistr} display the distribution of the parameters of the spectral models. $\Gamma_1$ is always steep, and can be interpreted as the hard tail of the accretion disk or the X-ray corona. $\Gamma$ and $\Gamma_2$ can be quite hard, clearly due to the jet dominance. A rough estimate the minimum flux needed to be detected by the present set of satellites and exposures is $\sim 10^{-12}$~erg~cm$^{-2}$~s$^{-1}$ (please remind that I have given priority to spectral variability, and, therefore, I have divided long exposures into hourly time bins). The spectrum observed by \emph{XMM-Newton} on August 18, 2018, $19:15:51$~UT was fitted by a broken power-law model, with $\Gamma_1=2.31_{-0.06}^{+0.05}$, $\Gamma_2=1.76_{-0.28}^{+0.20}$, $E_{\rm break}=2.9_{-0.4}^{+1.0}$~keV, $F_{1}\sim 8.43\times 10^{-12}$~erg~cm$^{-2}$~s$^{-1}$ and $F_{2}\sim 1.50\times 10^{-12}$~erg~cm$^{-2}$~s$^{-1}$. This can be considered the minimum observed jet contribution. The minimum disk contribution was recorded by \emph{Chandra} on November 8, 2021, $02:19:25$~UT. The parameters of the broken power-law model are: $\Gamma_1=2.66_{-0.66}^{+1.02}$, $\Gamma_2=1.56_{-0.25}^{+0.17}$, $E_{\rm break}=1.73_{-0.37}^{+0.85}$~keV, $F_{1}\sim 2.54\times 10^{-12}$~erg~cm$^{-2}$~s$^{-1}$ and $F_{2}\sim 7.16\times 10^{-12}$~erg~cm$^{-2}$~s$^{-1}$.

It is worth noting the presence of some cases of broken power-law with both photon indexes quite soft ($\Gamma \gtrsim 2$). These cases could be interpreted as the hard tail of the accretion disk ($\Gamma_1$) plus the corona dominating above $E_{\rm break}$, while the jet has negligible contribution. Another possibility seems to be an expedient when an absorption edge cannot fit well, because of small optical depth and a noisy or low-statistics spectrum. 

Fig.~\ref{edgesdistrib} displays the distribution of the rest frame energy of the absorption edges and their optical depths. The energies are basically concentrated around $\sim 0.3$ and $1.2-1.4$~keV, with some tail around $1.8$~keV. The low energy detections might be artifacts, since the low-energy threshold of the detectors is indeed $0.3$~keV: although the edges are generally blueshifted, some doubts remain. The $1.2-1.4$~keV edges seem to be more robust, also because such features have already been detected in other NLS1s (e.g. \citealt{LEIGHLY1997,KOMOSSA1999,REEVES2013}) and different interpretations have been put forward, including highly blueshifted OVII and/or OVIII ($v\sim 0.38-0.89c$ in the present case, \citealt{LEIGHLY1997}) or Fe L-shell absorption lines between $0.8-1$~keV rest frame, \citep{KOMOSSA1999,REEVES2013}, or Mg K-shell ($1.3$~keV rest frame, \citealt{REEVES2013}). The Fe L-shell interpretation is interesting, given the significant presence of iron in NLS1s. The $\sim 1.8$~keV is generally associated with Si VIII-IX \citep{REEVES2013}. Most of these edges are quite shallow ($\tau\lesssim 1$), but there are also many observations with moderate depth. 

\section{Final Remarks}
The hourly timescale of appearance/disappearance of the absorption edges implies a very small region. Assuming a spherical shape, then the radius of the region is: 

\begin{equation}
r<\frac{c\Delta t}{(1+z)}\sim 10^{14}\,\mathrm{cm}\sim 31r_{\rm g} 
\end{equation}

\noindent where $r_{\rm g}=3.2\times 10^{12}$~cm is the gravitational radius of the central black hole in 1H~$0323+342$. By assuming the wind speed measured by \cite{MEHDIPOUR2019}, $v_{\rm out}\sim 855$~km/s (average of the two values), then the distance from the central black hole is:

\begin{equation}
R>\frac{2GM}{v_{\rm out}^2}\sim 7.3\times 10^{15}\, \mathrm{cm}\sim 2281r_{\rm g}
\end{equation}

\noindent which satisfies the condition $r/R\lesssim 1$ \cite{BLUSTIN2005}. This places the absorber at the outer border of the accretion disk. The temptation to interpret this wind in light of the drag of mass from the disk to the jet according to \cite{BP1982} is quite strong. Indeed, it is expected that the \cite{BZ1977} theory alone could not explain the measured jet power of NLS1s \citep{FOSCHINI2011}.  

I would like to stress that this work is very preliminary, and a more detailed analysis of the spectral fitting and variability is ongoing.

\section*{Acknowledgements}
I would like to thank the director of Brera Astronomical Observatory of INAF, Roberto Della Ceca, for funding the travel to the conference.


\begin{thebibliography}{}
\bibitem[Abdo et al. (2009)]{LAT2009} Abdo, A.~A., et al., 2009, ApJ 707, L142.

\bibitem[Angelakis et al. (2015)]{ANGELAKIS2015} E. Angelakis, et al., 2015, A\&A 575, A55.

\bibitem[Berton et al. (2016)]{BERTON2016} M. Berton, et al., 2016, A\&A 591, A88.

\bibitem[Berton et al. (2020)]{BERTON2020} M. Berton, et al., 2020, A\&A 636, A64.

\bibitem[Blandford \& Payne (1982)]{BP1982} R.~D. Blandford, \& D.~G. Payne, 1982, MNRAS 199, 883.
\bibitem[Blandford \& Znajek (1977)]{BZ1977} R.~D. Blandford, \& R.~L. Znajek, 1977, MNRAS 179, 433.

\bibitem[Blustin et al. (2005)]{BLUSTIN2005} A.~J. Blustin, et al., 2005, A\&A 431, 111.

\bibitem[Cash (1979)]{CASH1979} W. Cash, 1979, ApJ 228, 939.

\bibitem[Czerny et al. (2009)]{CZERNY2009} B. Czerny, et al., 2009, ApJ 698, 840.

\bibitem[Evans et al. (2014)]{EVANS2014} P.~A. Evans, et al., 2014, ApJS 210, 8

\bibitem[Foschini et al. (2009)]{FOSCHINI2009} L. Foschini, et al., 2009, Adv. Space Res. 43, 889.

\bibitem[Foschini (2011)]{FOSCHINI2011} L. Foschini, 2011, RAA 11, 1266.

\bibitem[Foschini (2012)]{FOSCHINI2012} L. Foschini, 2012, in: ``Nuclei of Seyfert galaxies and QSOs - Central engine \& conditions of star formation'', November 6-8, 2012, Bonn (Germany), Proceedings of Science, Seyfert 2012, id 10. 

\bibitem[Foschini et al. (2015)]{FOSCHINI2015} L. Foschini, et al., 2015, A\&A 575, A13.

\bibitem[Foschini et al. (2019)]{FOSCHINI2019} L. Foschini, et al., 2019, Universe 5, 199.

\bibitem[HI4PI Collaboration (2016)]{HI4PI} HI4PI Collaboration, N. Ben Bekhti, L. Floer, et al., 2016, A\&A 594, A116. 

\bibitem[J\"arvel\"a et al. (2021)]{JARVELA2021} E. J\"arvel\"a, et al., 2021, Front. Astron. Space. Res. 8, 735310.

\bibitem[Komossa (1999)]{KOMOSSA1999} S. Komossa, 1999, In: ``ASCA/ROSAT Workshop on AGN and the X-ray Background'', November 1-3, 1999, Tokyo (Japan), edited by T. Takahashi and H. Inoue, ISAS Report, p. 149.

\bibitem[L\"ahteenm\"aki et al. (2018)]{LAHTEENMAKI2018} A. L\"ahteenm\"aki, et al., 2018, A\&A 614, L1.

\bibitem[Landt et al. (2017)]{LANDT2017} H. Landt, et al., 2017, MNRAS 464, 2565.

\bibitem[Leighly et al. (1997)]{LEIGHLY1997} K. Leighly, et al., 1997, ApJ 489, L25.

\bibitem[Longo et al. (2025)]{LONGO2025} F. Longo, et al., 2025, ATel 17407.

\bibitem[Mehdipour \& Costantini (2019)]{MEHDIPOUR2019} M. Mehdipour \& E. Costantini, 2019, A\&A 625, A25.

\bibitem[Paliya et al. (2014)]{PALIYA2014} V.~S. Paliya, et al., 2014, ApJ 789 143.

\bibitem[Protassov et al. (2002)]{PROTASSOV2002} R. Protassov, et al., 2002, ApJ 571, 545.

\bibitem[Reeves et al. (2013)]{REEVES2013} J.~N. Reeves, et al., 2013, ApJ 776, 99.

\bibitem[Riess et al. (2022)]{RIESS2022} A.~G. Riess, et al., 2022, ApJ 934, L7.

\bibitem[Romano et al. (2023)]{ROMANO2023} P. Romano, et al., 2023, A\&A 673, A85.

\bibitem[Rosa et al. (2025)]{ROSA2025} V. Rosa, et al., 2025, A\&A 698, A160.

\bibitem[Tibolla et al. (2013)]{TIBOLLA2013} O. Tibolla, et al., 2013, in: ``33rd International Cosmic Ray Conference'', July 2-9, 2013, Rio de Janeiro, Brazil. Edited by Alberto Saa, p.2748. 

\bibitem[Wilms et al. (2000)]{WILMS2000} J. Wilms, et al., 2000, ApJ 542, 914. 

\bibitem[Yao et al. (2023)]{YAO2023} S. Yao, \& S. Komossa, 2023, MNRAS 523, 441.

\bibitem[Zhou et al. (2007)]{ZHOU2007} H. Zhou, et al., 2007, ApJ 658, L13.

\end{thebibliography}
\end{document}